\documentclass[%
 reprint,
 amsmath,amssymb,
 aps,
]{revtex4-2}

\pdfoutput=1

\usepackage{float}
\usepackage{subcaption}
\usepackage{notoccite}
\usepackage{caption}
\usepackage{comment}
\usepackage{enumitem}
\usepackage{graphicx}
\usepackage{dcolumn}
\usepackage{bm}
\usepackage[a4paper, left=1.5cm, right=1.5cm, top=3cm, bottom=3cm]{geometry} 

\usepackage[breaklinks,pdftex,colorlinks=true,allcolors=blue]{hyperref}
\usepackage{tikz}
\usepackage{multirow}
\usepackage{xcolor, lipsum}

\usepackage{indentfirst}

\usepackage[utf8]{inputenc}

\begin{document}

\title{\Large{Alfvénic high-frequency oscillations at the pedestal of JET plasmas}}

\author{L. Roque$^1$, P. Rodrigues$^1$, E. R. Solano$^2$, JET contributors$^a$ and The Eurofusion Tokamak Exploitation Team$^b$ \\ \bigskip $^1$ Instituto de Plasmas e Fusão Nuclear, Instituto Superior Técnico, Universidade de Lisboa, 1049-001 Lisboa, Portugal\\
$^2$ Laboratorio Nacional de Fusion, CIEMAT, 28040 Madrid, Spain \\ $^a$See the author list of C.F. Maggi \textit{et al.} Nucl. Fusion \textbf{64} 112012 (2024) \\ $^b$See the author list of  N. Vianello \textit{et al.} Nucl. Fusion \textbf{66} 116010 (2026)}

\date{September 22, 2026}

\begin{abstract}

Long-lived ($\sim 10$ s) high-frequency oscillations (HFOs, $50-450$ kHz) near the plasma edge have been reported and experimentally described in L-H transition studies at JET and AUG. In this work, we show that these HFOs are a general phenomenon observed in various plasma scenarios and compositions (H$^1$, D, D-He$^3$, D-T) under different heating schemes, including pure Ohmic, ICRH and NBI. We demonstrate that dominant axisymmetric ($n=0$) HFOs are Global Alfvén eigenmodes (GAEs) tied to the shear-Alfvén continuum (SAC) minima arising at the plasma edge due to the sharp decrease of the density and increase of the safety factor. This hypothesis is corroborated by the remarkable agreement found between the measured frequencies of HFOs and the SAC minima computed by the ideal MHD code \texttt{CSMISH} for a comprehensive set of JET pulses. This result, along with the fact that HFOs are observed over long time windows and across a variety of plasma scenarios, makes them convenient MHD constraints for accurate equilibrium reconstruction near the edge. In view of this pragmatic application, we derive a first-order analytic expression for the two lowest coupled branches of the $n=0$ SAC, disclosing their dependence (and, consequently, that of the HFOs) on the plasma density, safety factor, and elongation profiles. In addition, we discuss the nature of $n\neq 0$ HFOs observed along with the dominant $n=0$ HFOs.

\end{abstract}

\maketitle

\section{Introduction}

Long-lived quasi-coherent high-frequency magnetic oscillations at the plasma edge have been reported close to the L-H transition at JET \cite{Vianello2015, Refy2020} and AUG \cite{Refy2020} tokamaks. The dominant oscillations are found to have toroidal mode number $n=0$ and poloidal mode number $m=1$~\cite{Vianello2015}. 
These axisymmetric ($n=0$) HFOs are measured in the spectrogram of Mirnov coils during both low and high confinement modes (although attenuated in the latter), and also during a transition or intermediary phase with better confinement than the L-mode but not yet a fully developed H-mode, known as M-mode at JET \cite{Solano2017} and I-phase at AUG \cite{Birkenmeier2016}. This phase is characterised by improved confinement properties, marked by the formation of a weak temperature pedestal and steepening of the plasma density pedestal, and is accompanied by $n=0$, $m=1$ (up-down) low-frequency coherent and multi-harmonic magnetic oscillations (LFOs). These LFOs have a fundamental harmonic of order $0.5-2$ kHz and affect various local measurements across the pedestal and scrape-off layer regions \cite{Vianello2015, Solano2017, Refy2020, Birkenmeier2015, Birkenmeier2016}. During the M-mode, HFOs and LFOs exhibit similar temporal evolution and were proven to be correlated \cite{Vianello2015, Refy2020}. While the frequencies of LFOs measured at JET were found to scale with the edge poloidal Alfvén speed in RF heated plasmas with slow power ramps \cite{Solano2017}, no conclusive explanation concerning the nature of HFOs, supported by a convincing frequency scaling law, has been provided. So far, an analytic model based on the drift-Alfvén model proposed that such oscillations could be kinetic Alfvén waves \cite{Grover2024}. However, the model's predictions overestimate by a factor of 2 the experimental measurements performed at several devices (among which JET and AUG). 

Although initially reported in L-H transition studies, we found that HFOs at JET are not restricted to such scenarios and appear in a variety of plasma compositions (H$^1$, D, D-He$^3$, D-T) with a wide range of plasma parameters and under different heating schemes, namely pure Ohmic, neutral beam injection (NBI) and ion cyclotron resonance heating (ICRH). The frequencies of the two lowest HFOs invariably lie within the Toroidicity-induced Alfvén Eigenmode (TAE) and Ellipticity-induced Alfvén Eigenmode (EAE) frequency range.

We intend to demonstrate that these broadband axisymmetric HFOs can be explained by ideal MHD theory as $n=~0$ Global Alfvén Eigenmodes (GAEs), that is, discrete modes of plasma oscillation that have been described in earlier publications for general toroidal mode numbers \cite{Appert1982} and for the particular case of $n=0$ \cite{Villard1997}. To this end, the approach adopted here was to compare the measured frequencies of the $n=0$ HFOs with those expected for $n=0$ GAEs as estimated from the minima of the incompressible continuous Alfvén spectra computed by the ideal MHD code \texttt{CSMISH}~\cite{Huysmans2001} for a diverse set of JET pulses covering a wide range of plasma parameters. Besides clarifying the nature of these oscillations, this work also aims to provide a predictive model for their frequencies and point out their practical applications. This is of interest not only to JET, but also to present and future fusion devices, since the excitation of quasi-coherent axisymmetric GAEs has been previously reported at MAST \cite{McClements2002} and COMPASS \cite{Markovic2017} for Ohmic deuterium plasmas, and inferred for TFTR, which presented a similar Alfvénic activity~\cite{chang1995}. At JET, only coherent $n=0$ GAEs have been discussed, and exclusively in the context of destabilisation by energetic particles arising either from external heating (ICRH/NBI) \cite{Oliver2017} or fusion-born alpha particles \cite{Oliver2026}. In contrast, the broadband, long-lived $n=0$ GAEs reported herein are found across a wider range of conditions (including in the absence of energetic particles) and are analogous to those observed at MAST and COMPASS.

This paper is organised as follows. In section \ref{sec:Experimental_Description_HFOs}, we present and discuss the main properties of the $n=0$ HFOs using two pulses as examples. Section \ref{sec:GAEs_SAC} gives a brief overview of axisymmetric GAEs in light of ideal MHD and revises the coupling model of the incompressible continuous spectra. In order to understand how the frequencies of $n=0$ GAEs relate to plasma shaping parameters, we derive, in section \ref{sec:AnalyticalSAC}, a first-order analytic expression for the two lowest coupled branches of the shear-Alfvén continuum. The purpose of this derivation is justified in section \ref{sec:Numerical_GAEs}, where we demonstrate numerically that the frequencies of $n=0$ continuum minima are a good proxy for the frequencies of $n=0$ GAEs. This finding is followed by the main results of this work, presented in section \ref{sec:Extended_analysis}, where we show the remarkable agreement between the frequencies of the $n=0$ HFOs and those of the $n=0$ shear-Alfvén continuum minima. The limitations of ideal MHD codes like \texttt{MISHKA} \cite{Mikhailovskii1997} to compute the GAEs' structure are addressed in section \ref{sec:Resistivity}, as well as the role of non-ideal terms, in particular the plasma resistivity. Section \ref{sec:HFOs_of_finite_n} dedicates some brief words to non-axisymmetric ($n \neq 0$) HFOs observed together with the dominant $n=0$ HFOs. The main conclusions and prospects of this work are then summarised in section \ref{sec:conclusions}.

\section{Experimental overview of axisymmetric HFOs}\label{sec:Experimental_Description_HFOs}

Two $n=0$ high-frequency quasi-coherent bands of medium to low amplitude can be observed in the spectrogram of the poloidal magnetic signal $\dot{B}_\theta$ measured by Mirnov coils, and, in particular, in the processed phase spectrogram displaying the toroidal mode numbers estimated from the combined analysis of the signal of all available coils. At JET (major radius $R_0 \approx 3$ m and minor radius $a \approx 0.9$ m), the frequency of the first band typically ranges between $50-150$ kHz, while the second ranges between $100-450$ kHz, depending on the plasma scenario and parameters. The lower-frequency oscillation is generally broader and more intense than the higher-frequency one (hereafter referred to as HFO-1 and HFO-2, respectively). The latter, being weaker, is not always obvious in the spectrogram of $\dot{B}_\theta$, and both bands are generally more pronounced in the signal of coils located at the low-field side. In some cases, a third and even fourth band are detected as well. All these $n=0$ HFOs are observed already before external heating power is turned on, remain visible during the heated phase, and prevail after auxiliary heating is turned off. Repeatedly, additional HFOs bands of finite toroidal mode number ($n\neq0$) appear along with $n=0$ HFOs, exhibiting an identical modulation. Their main properties are described next, employing a sample of two pulses from the set of 14 that was analysed in this work: \#80951 and \#82221, from the L-H transition studies \cite{Solano2017, Solano2013}.

\begin{figure}[t!]
    \centering
    \includegraphics[width=\linewidth]{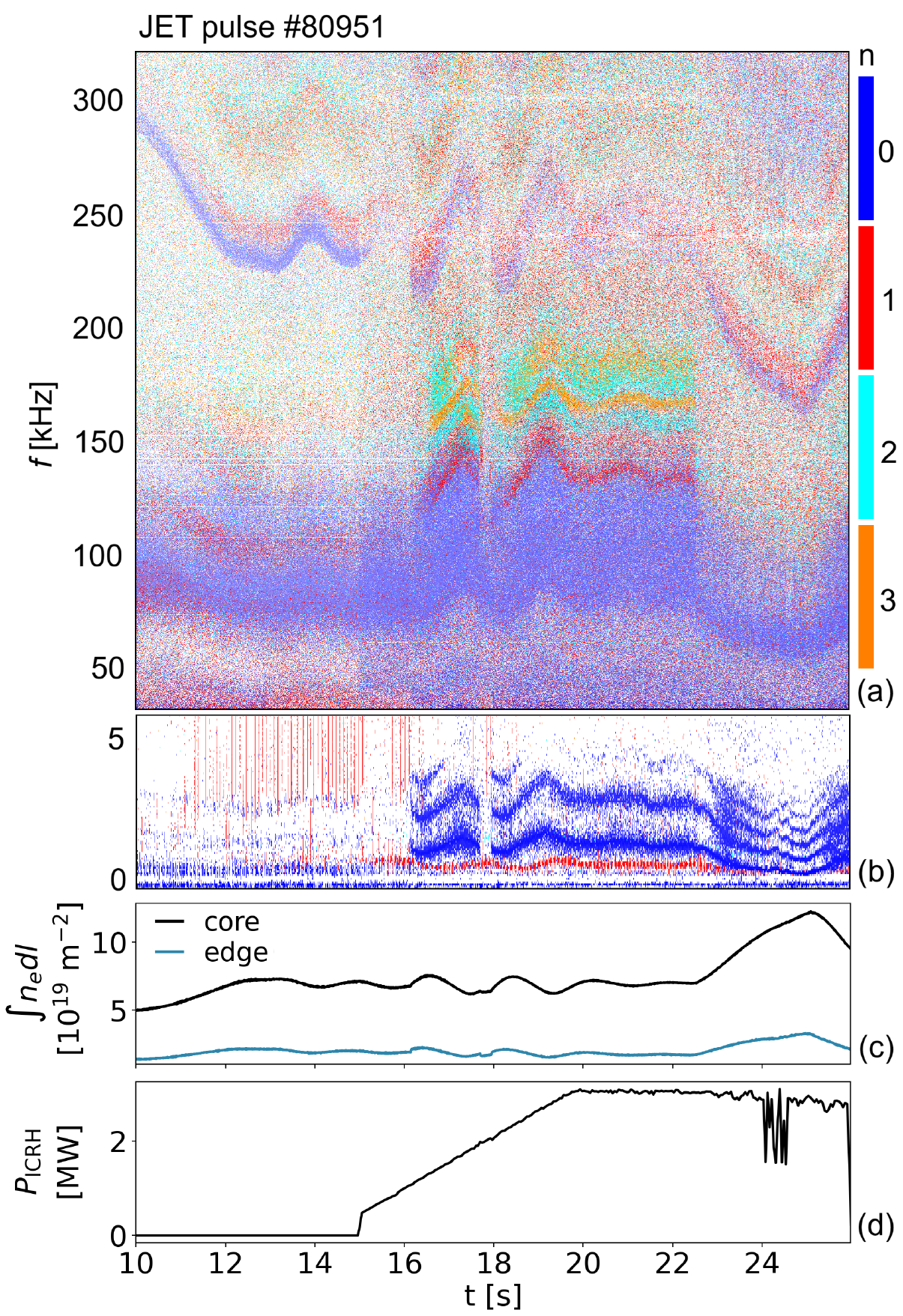}
    \caption{Toroidal mode number spectrogram in the (a) high and (b) low frequency range; (c) core and edge line integrated electron density from interferometer measurements; (d) ICRH heating power for JET pulse \#80951.}
    \label{fig:shot_80951}
\end{figure}

\begin{figure}[h!]
	\centering
	\includegraphics[width=\linewidth]{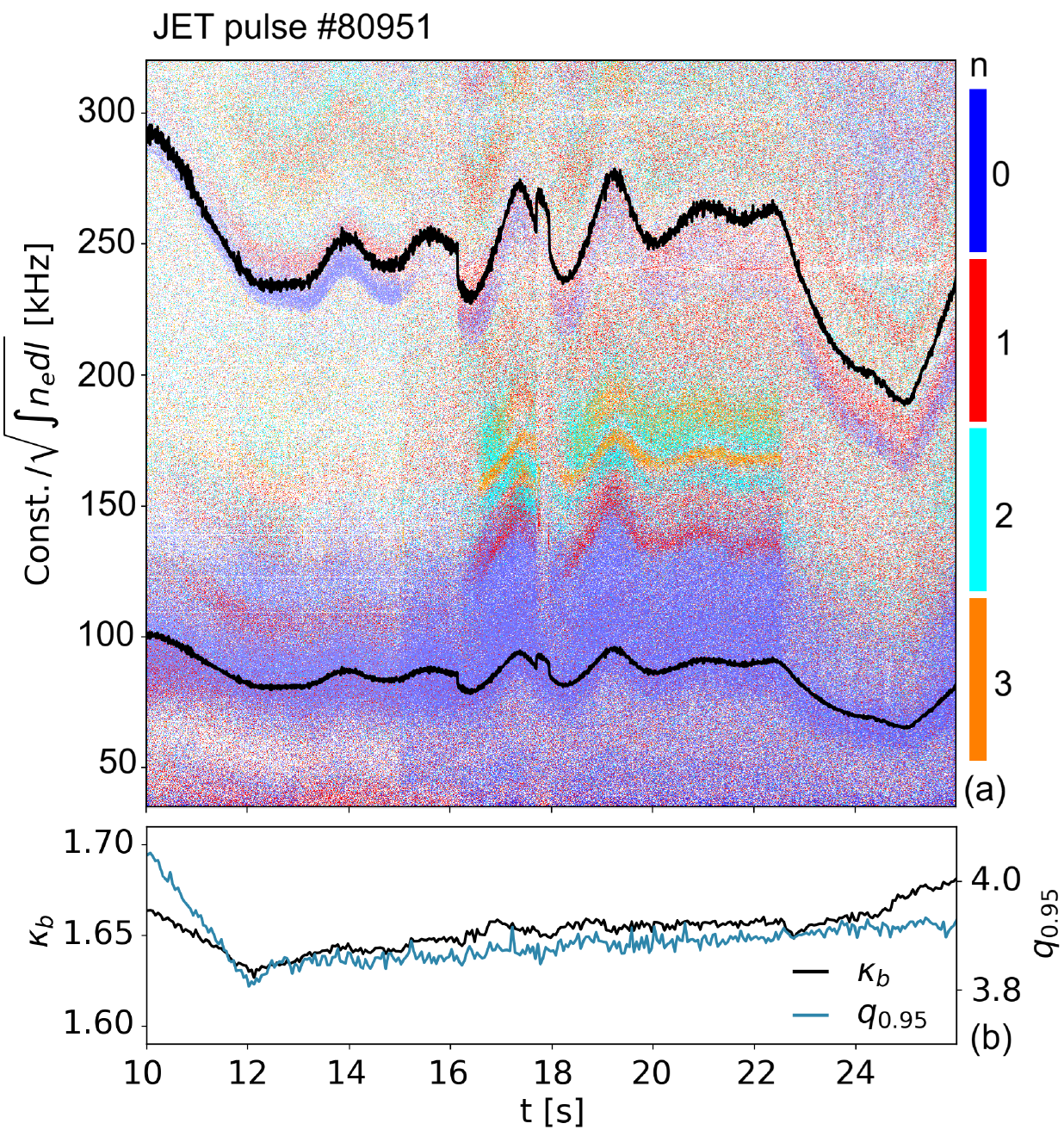}
	\caption{(a) Inverse square root of the edge line integrated electron density from interferometry measurements, normalised to the frequency value of each HFO band at t = 10 s (100 kHz and 290 kHz) superposed to the toroidal mode number spectrogram; (b) temporal evolution of the elongation at the plasma boundary and safety factor at $\Psi_n = 0.95 $, with $\Psi_n$ the poloidal magnetic field normalised to its value at the boundary, for JET pulse \#80951.}
	\label{fig:discharge_80951_plots_inv_sqrt_LID4_mult_factor}
\end{figure}

\begin{figure}[h!]
	\centering
	\includegraphics[width=\linewidth]{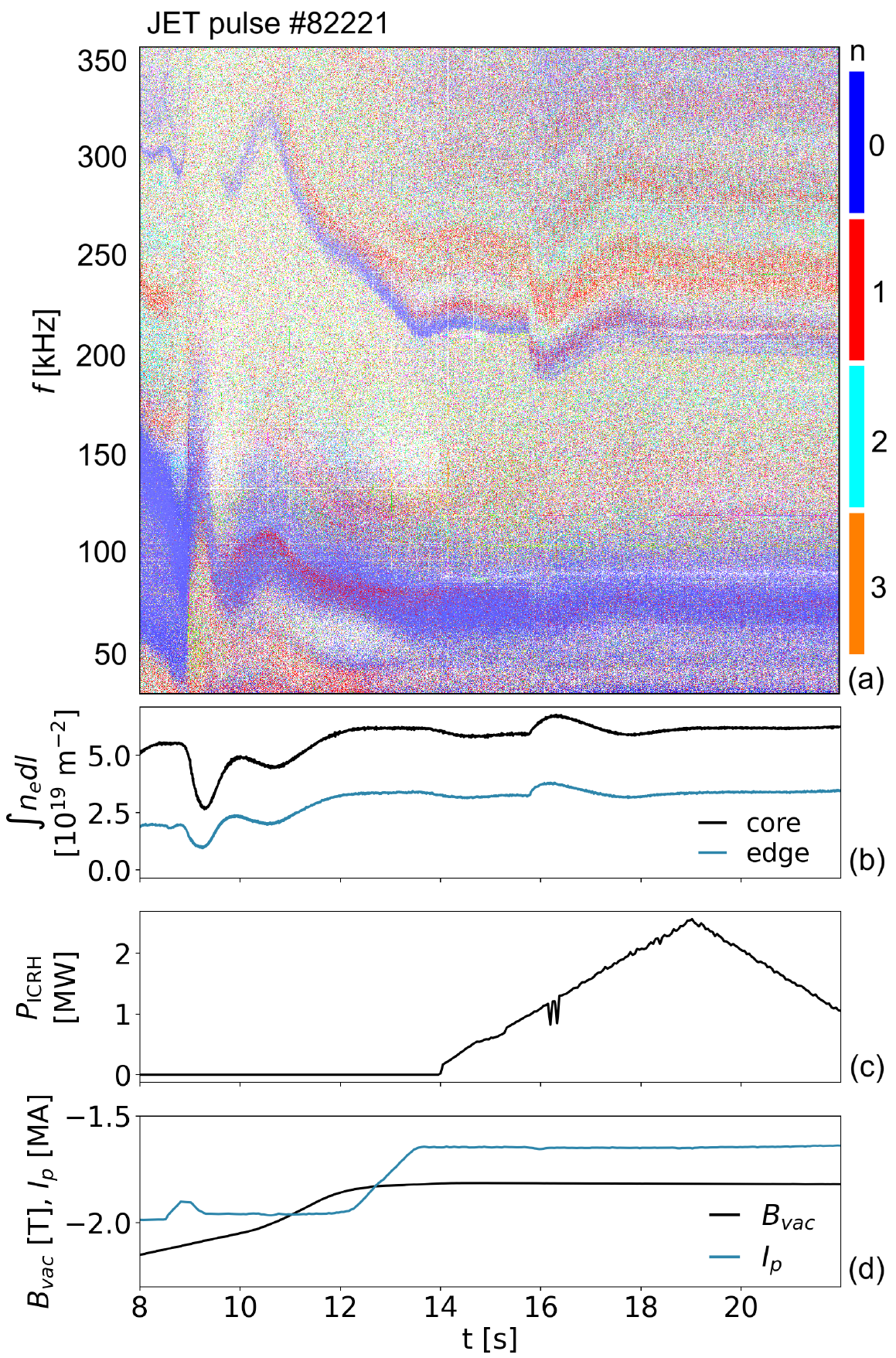}
	\caption{(a) Toroidal mode number spectrogram in the high frequency range; (b) core and edge line integrated electron density from interferometer measurements; (c) ICRH heating power; (d) temporal evolution of $B_\text{vac}$ and $I_p$ for JET pulse \#82221.}
	\label{fig:discharge_82221_plots_spec}
\end{figure}

Pulse \#80951 is a deuterium plasma heated by ICRH with vacuum toroidal magnetic field at the geometric centre $B_\text{vac} = 2.4$ T and plasma current $I_p = 2.0$ MA. The toroidal mode number spectrogram for low and high frequency ranges, along with the line integrated electron density taken from interferometry measurements and ICRH power, is presented in figure \ref{fig:shot_80951}. Both the HFO-1 ($80-140$ kHz) and HFO-2 ($160-300$ kHz) are modulated by the electron density (in fact, $1/\sqrt{n_e}$, with $n_e$ the electron density) and, as previously reported \cite{Vianello2015, Refy2020}, exhibit a temporal evolution similar to the LFOs, with frequencies in the range $1-4$ kHz. Although the lower-frequency HFO band appears to have a milder frequency modulation than the upper one, figure \ref{fig:discharge_80951_plots_inv_sqrt_LID4_mult_factor} shows that, apart from a multiplicative factor (which will depend on other plasma parameters), the modulation is indeed the same. The clear frequency dependence on the inverse square root of the edge electron density, which, under quasi-neutrality, is approximately proportional to the plasma mass density, hints at the Alfvénic nature of HFOs. For this pulse, ICRH is turned on at $t=15$ s and turned off at $t=26$ s. The $n=0$ HFOs are already present before that, when the plasma is still in its Ohmic heating phase.

Pulse \#82221 is a deuterium plasma as well, again heated by ICRH. In this case, both toroidal field and plasma current evolve in time as shown in figure \ref{fig:discharge_82221_plots_spec}(d). Just as in the previous example, the HFOs are visible before external heating power starts (at $t=14$ s). In addition to the $n=0$ bands, low-amplitude $n=1$ bands above or coincident to $n=0$ ones appear in the toroidal mode number spectrogram, as shown in figure \ref{fig:discharge_82221_plots_spec}(a).  These additional oscillations are seen in multiple JET pulses, both during Ohmic plasma phase and when external heating mechanisms are on, and their modulation with the $n=0$ HFOs suggests an identical Alfvénic nature, as will be discussed further on.

\section{Axisymmetric Shear-Alfvén continuum and Global Alfvén Eigenmodes}\label{sec:GAEs_SAC}

Ideal MHD theory predicts the existence of discrete Alfvén Eigenmodes (AEs) at the frequency gaps induced in the continuous Alfvén spectrum by the geometry of the equilibrium magnetic field (toroidicty, elongation, etc \cite{Cheng1986,Betti1991, Heidbrink2008}) and/or around radial positions where the continuum reaches an extremum in response to plasma density and current density gradients \cite{Appert1982, Villard1997}. Axisymmetric shear AEs necessarily require the last condition to be met in order to exist due to the particular properties of the $n=0$ shear-Alfvén continuous spectrum, which shall be briefly clarified below.

Under the usual small-perturbation formulation of the plasma displacement, $\bm{\xi}(\bm{x},t)=\bm{\xi}(\bm{x}) e^{-i \omega t} := \bm{\xi}e^{-i \omega t}$, the linearised set of ideal MHD equations reduces to the force balance equation \cite{Bernstein1958}
\begin{equation}\label{eq:forceBalance}
F(\bm{\xi}) = -\mu_0 \rho \omega^2 \bm{\xi},
\end{equation}

\noindent
where $\rho$ is the plasma mass density, $\mu_0$ is the magnetic constant, and  $F(\bm{\xi}) = (\nabla \times \bm{B}) \times \bigr[\nabla \times (\bm{\xi} \times \bm{B})\bigr] + \bigr[\nabla \times\nabla \times (\bm{\xi} \times \bm{B})\bigr] \times \bm{B} +\mu_0 \nabla(\gamma P \nabla \cdot\bm{\xi} + \bm{\xi}\cdot \nabla P)$ is the ideal MHD operator, with $\gamma$ the adiabatic constant, $P$ and $\bm{B}$ the equilibrium pressure and magnetic field. For each radial position, there is a set of $\omega^2$ values that turn $\bm{\xi}$ singular due to vanishing coefficients in the force balance equation, overall forming a continuous spectrum of singular solutions (i.e. the Alfvén continuum). Waves with frequencies in the continuum are strongly damped and, thus, cannot be easily driven in the plasma. However, a discrete spectrum of AEs can be found outside the continuum.

Rewriting the force balance equation \eqref{eq:forceBalance} as a matrix problem \cite{Cheng1986} and discarding all compressible contributions (i.e. $\nabla \cdot \boldsymbol{\xi}=~0$), the shear-Alfvén continuum (SAC) is formed by the set of $\omega^2$ values that yield non-trivial solutions for equation

\begin{equation}\label{eq:EigenvalueEq}
	\Biggr[\frac{\omega^2 \mu_0 \rho |\nabla \Psi|^2}{B^2}+ \boldsymbol{B} \cdot \nabla \left(\frac{|\nabla \Psi|^2}{B^2} \boldsymbol{B} \cdot \nabla \right) \Biggr] \xi^A = 0 ,
\end{equation}

\noindent
where $\xi^A = \boldsymbol{\xi} \cdot \boldsymbol{B} \times \nabla \Psi /|\nabla \Psi|^2 $ is the shear-Alfvén (or binormal) component of the plasma displacement.

In the cylindrical limit, the SAC reduces to \cite{Appert1974}

\begin{equation}\label{eq:SACcylind}
	\omega^2 = k_\parallel^2 v_A^2 \approx \frac{1}{R^2}\left(n-\frac{m}{q}\right)^2 v_A^2,
\end{equation}

\noindent
with $q$ the safety factor, $R$ the distance to the torus axis, $k_\parallel \approx (n - m/q)/R$ the wave vector parallel to the magnetic field and $v_A = B/\sqrt{\mu_0 \rho } $ the Alfvén speed. For the special case of axisymmetric perturbations ($n=0$), equation \eqref{eq:SACcylind} becomes simply $\omega^2 = m^2 v_A^2/(q^2R^2) \approx m^2  v_{A,\text{pol}}^2 /r^2$ ($r$ being the distance to the magnetic axis), implying that continuum branches corresponding to Fourier harmonics with different $|m|$ will never cross \cite{Villard1997}. This peculiarity extends from cylindrical to toroidal geometry, where no frequency gaps are produced exclusively by coupling axisymmetric harmonics with distinct $|m|$. Therefore, we only expect discrete $n=0$ shear AEs to appear near continuum extrema (which are accumulation points of discrete modes) produced by radial profiles. These are known as Global Alfvén Eigenmodes, as they generally extend from the core to the plasma edge \cite{Appert1982, Villard1997}.

\begin{figure}[t!]
	\centering
	\includegraphics[width=\linewidth]{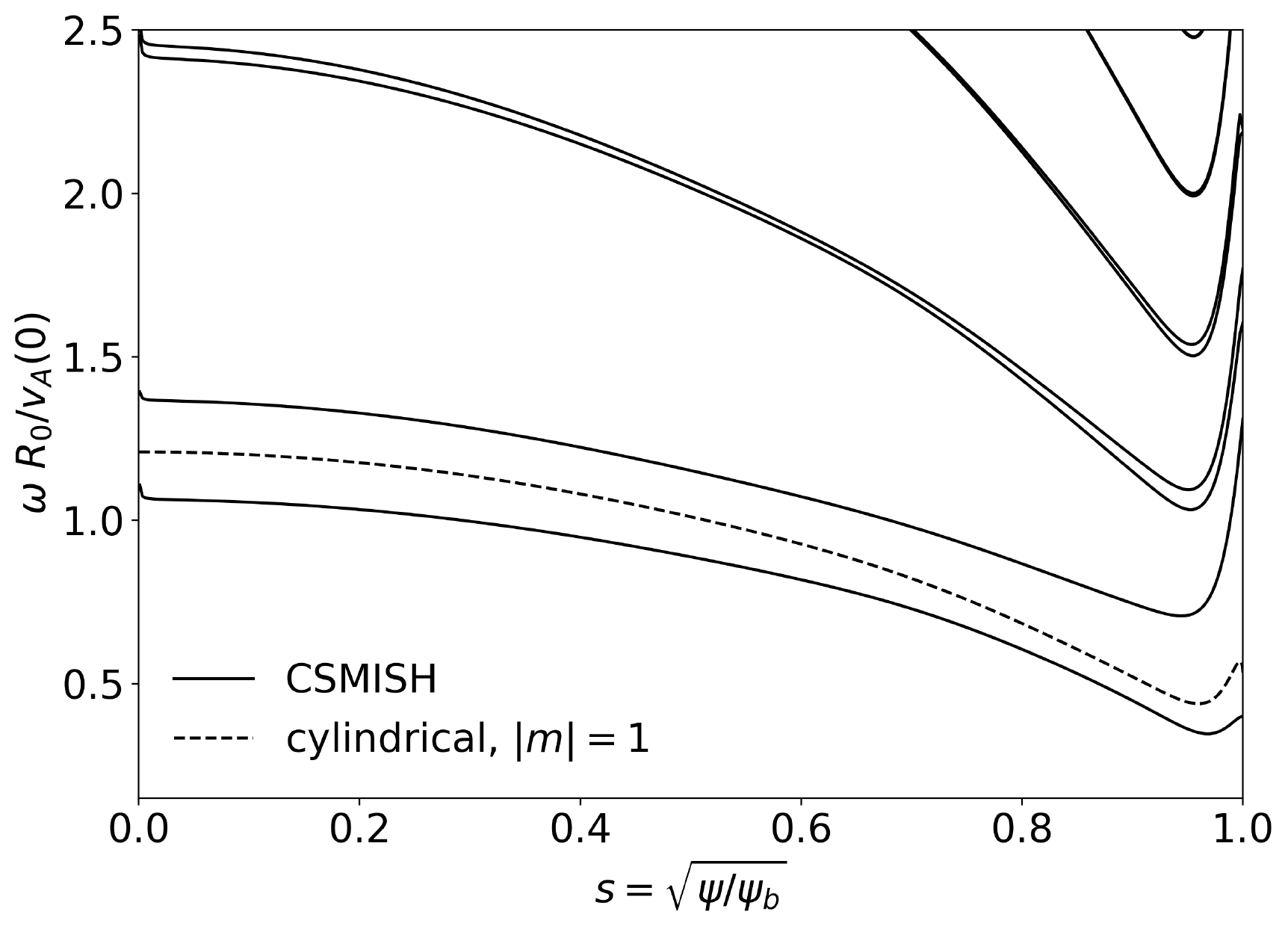}
	\caption{Normalised $n=0$ SAC computed by \texttt{CSMISH} for pulse \#82221 at $t=12.47$ s, along with the cylindrical SAC calculated from equation \eqref{eq:SACcylind} by setting $R \approx R_0$, in terms of the radial coordinate $s = \sqrt{\Psi/\Psi_b}$, with $\Psi/\Psi_b$ the poloidal field flux normalised to its value at the boundary.}
	\label{fig:SAC_t5247_cylindrical}
	\vspace{-2mm}
\end{figure}

While frequency branches with different $|m|$ do not cross in the $n=0$ SAC, those with the same $|m|$ overlap and their harmonics become coupled in equation \eqref{eq:EigenvalueEq}, therefore opening a continuous frequency gap that extends from the core to the plasma edge (see figure \ref{fig:SAC_t5247_cylindrical}).
In particular, the size of the frequency gap squared, $\Delta \omega^2$, arising from the coupling of the harmonics $m=+1$ and $m=-1$ was numerically shown to increase linearly with the plasma ellipticity in the incompressible regime \cite{Oliver2017}. Although ellipticity is not the only factor contributing to the emergence of this gap, another being the geodesic curvature whose effects are introduced when compressible terms are included \cite{zhou2020}, it is by far the most relevant of the two. This statement is supported by the nearly exact match (not shown here) of the shear branches coming from the continuum computed by \texttt{CSCAS} \cite{Poedts1993} in the compressible ($\gamma=5/3$) and incompressible ($\gamma \rightarrow 0$) regimes.

Near the plasma edge, the decrease in density, combined with the sharp increase in $q$, gives rise to a minimum in each frequency branch of the SAC, as illustrated in figure \ref{fig:SAC_t5247_cylindrical}. The radial position of the minimum of the lower and upper coupled $|m|=1$ branches is not the same, although both rest near the top of the pedestal: the lower minimum is located near the region of sharp density gradient variation, whereas the upper one is shifted to the inner side of the pedestal top. For $|m|\geq 2$, the coupling between branches is weaker, the frequency gap is smaller, and the radial position of the continuum minima almost coincide with one another and with the pedestal top.

Low-frequency GAEs, specifically those that lie below the minima of the coupled $|m|=1$ SAC branches, are the main focus of this work. As we will show later on, the frequencies of GAEs are tied to the SAC minima, which can hence accurately predict their values. Motivated by this, we derived an analytic expression for the two lowest SAC branches.

\subsection{Analytic coupling of the $|m|=1$ continuum branches} \label{sec:AnalyticalSAC}

We employed a local magnetic equilibrium model with up-down asymmetric cross section \cite{Rodrigues2018} to solve equation \eqref{eq:EigenvalueEq} in the coordinate system $(r, \theta, \phi)$ (with $\theta$ and $\phi$ the poloidal and toroidal angles, respectively, and $r$ the distance to the magnetic axis normalised to $a$) and derive an analytic expression for the lowest branches of the coupled SAC in terms of equilibrium shaping coefficients. 

To avoid unnecessary complexity, we simplified the original poloidal-field flux description \cite{Rodrigues2018} by retaining only first-order terms in the inverse aspect ratio, $\varepsilon = a/R_0$, such that

\begin{equation}
	\Psi(r, \theta) = \Psi_b S_0 r^2 \Bigr[ \Theta_0 (\theta) + \varepsilon r \Theta_1 (\theta) \Bigr],
\end{equation}

\noindent
with $\Psi_b$ the poloidal flux at the boundary, $S_0$ a geometric parameter that relates with the cylindrical $q$ and on-axis magnetic field, $B_0$, at lowest order as $S_0=\frac{a^2 B_0}{2 q \Psi_b}$, and the functions $\Theta_{0,1}(\theta)$ defined as follows:

\begin{equation}
	\begin{aligned}
		\Theta_0 (\theta) &= 1 +\kappa_s \cos 2\theta +\kappa_a \sin 2 \theta\\
		\Theta_1 (\theta) &= \Delta_s \cos \theta + \frac{1}{4} \kappa_a \sin\theta + \eta_s \cos 3 \theta + \eta_a \sin 3 \theta.
	\end{aligned}
\end{equation}

\noindent
The geometric coefficients $\Delta_s$, $\kappa_s$, and $\eta_s$ can be related at leading order with the conventional definitions \cite{Miller1998} of the Shafranov shift, $\Delta$, plasma elongation, $\kappa$, and triangularity, $\delta$, as \cite{Rodrigues2018}

\begin{equation}
	\begin{aligned}
		\Delta &\approx -\frac{a \varepsilon \Psi}{2 S_0} \frac{\Delta_s + \eta_s}{(1+\kappa_s)^2} , \quad \kappa \approx \sqrt{\frac{1+\kappa_s}{1-\kappa_s}}, \\
		&\delta \approx \varepsilon \sqrt{\frac{\Psi}{S_0(1+\kappa_s)}}\frac{\kappa_s(\Delta_s-\eta_s)-2\eta_s}{1-\kappa_s^2},
	\end{aligned}
\end{equation}
\noindent
while $\kappa_a$ and $\eta_a$ describe the up-down asymmetry of general tokamak cross sections. The ordering of each geometric constant matters to perform consistent series expansions in the small parameter $\varepsilon$. For typical JET equilibria, these generally follow $\Delta_s \sim 1$, $\kappa_s \sim \varepsilon$ and $\kappa_a , \eta_s , \eta_a \lesssim \varepsilon^2$ \cite{Rodrigues2018}.

To simplify the analysis and following previous approaches \cite{Oliver2017}, we considered that only the harmonics $m=-1$ and $m=1$ intervene for the first gap opening in the axisymmetric SAC by coupling with each other. Since $n=0$, all dependence on $\phi$ vanishes, such that the spatial part of the plasma displacement can be Fourier expanded as $\sum_{m} \boldsymbol{\xi}_m(\Psi (r, \theta)) e^{i m \theta}$. Therefore, $\xi^A  = \xi_{-1}^A e^{-i \theta}+\xi_{1}^A e^{i \theta}$ and equation \eqref{eq:EigenvalueEq} can be rewritten in matrix form as

\begin{equation}\label{eq:matrixFirstOrder}
\begin{bmatrix} e^{-i \theta} & e^{i \theta} \end{bmatrix}
\begin{bmatrix} \frac{(1+\tilde{\rho} \tilde{\omega}^2 q^2)\kappa_s}{2 q^2} & \tilde{\rho} \tilde{\omega}^2 -\frac{1}{q^2} \\ \tilde{\rho} \tilde{\omega}^2 -\frac{1}{q^2} & \frac{(1+\tilde{\rho} \tilde{\omega}^2 q^2)\kappa_s}{2 q^2} \end{bmatrix}
\begin{bmatrix}  \xi_{-1}^A \\  \xi_{1}^A \end{bmatrix}=0,
\end{equation}

\noindent
where $\tilde{\omega}=\omega R_0/v_A(0)$ and $\tilde{\rho}= \rho/\rho(0)$ are the frequency and mass density normalised to their on-axis values. Second-order terms in $\varepsilon$ and $\kappa_s$ were dropped in the entries of the matrix above.

The resulting expression for the two lowest branches of the coupled continuum will be given by the set of frequencies for which the determinant of the matrix vanishes. This condition is met when

\begin{equation}\label{eq:SACexpression}
	\tilde{\omega}_{\pm}^2 =   \frac{2 \pm \kappa_s}{\tilde{\rho} q^2 (2 \mp \kappa_s)}.
\end{equation}

To no surprise, in this first-order approach, we find that the factor responsible for the coupling of the $m=\pm1$ harmonics and the consequent separation of their corresponding frequency branches is the plasma ellipticity. Taking the continuum frequencies in equation \eqref{eq:SACexpression}, the size of the gap is given by 

\begin{equation}
	\tilde{\omega}_{+}^2-\tilde{\omega}_{-}^2 = \Delta \tilde{\omega}^2 = \frac{8 \kappa_s}{\tilde{\rho} q^2 (4-\kappa_s^2)}.
\end{equation}

\noindent
In the limit $\kappa_s^2 \ll~1$, the expression agrees with the linear relation between $\Delta \omega^2$ and $\kappa_s$ that was found numerically by scanning a range of the latter parameter \cite{Oliver2017}.

We must point out, however, that this simplistic approach may fail to capture the behaviour of the continuum at the edge due to large radial gradients of the geometric parameters, which have been neglected in the derivation. Improving the analytic expression to account for these effects will be the focus of a future work.

\section{Numerical modelling of axisymmetric HFOs and GAEs}\label{sec:HFOs_are_GAEs}

\subsection{Detailed analysis: an example of coupled SAC and discrete GAEs}\label{sec:Numerical_GAEs}

Next, we present the main properties of GAEs and their relationship with the continuum based on the numerical analysis performed on pulse \#82221 at $t=8.07$~s. For this specific time instant, the experimental frequencies of the two axisymmetric HFOs measured in the spectrogram of a Mirnov coil located at the low-field side (rather than in the toroidal mode number spectrogram) are $f_\text{HFO-1}=145 \pm 5$ kHz and $f_\text{HFO-2}=302 \pm 2$ kHz.

\begin{figure}[t!]
	\centering
	\includegraphics[width=\linewidth]{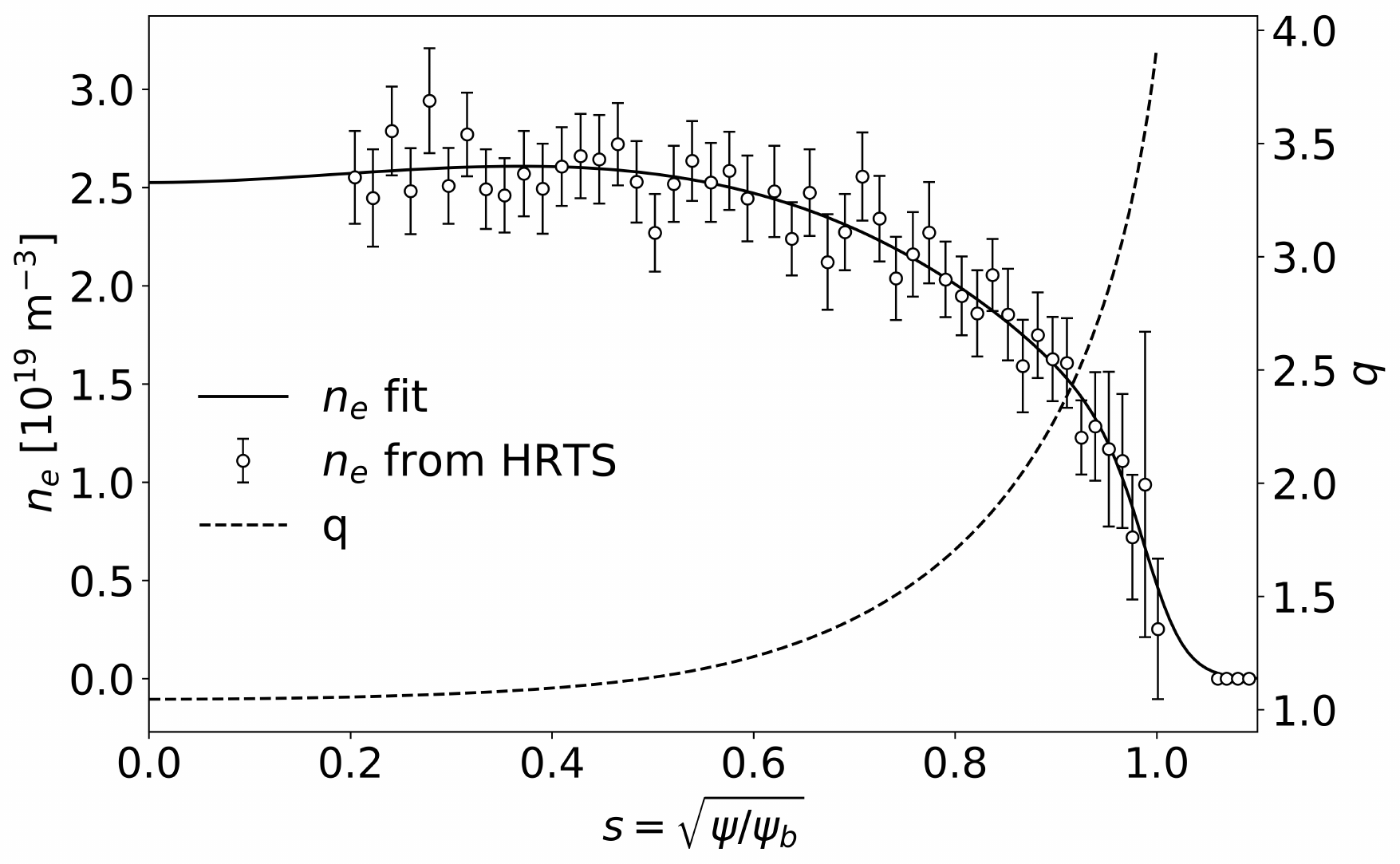}
	\caption{Fitted electron density, $n_e$, along with HRTS measurements and the $q$ profile computed by \texttt{HELENA} for pulse \#82221 at $t=8.07$ s.}
	\label{fig:ne_q_t4807}
\end{figure}

\begin{figure}[t!]
	\centering
	\includegraphics[width=\linewidth]{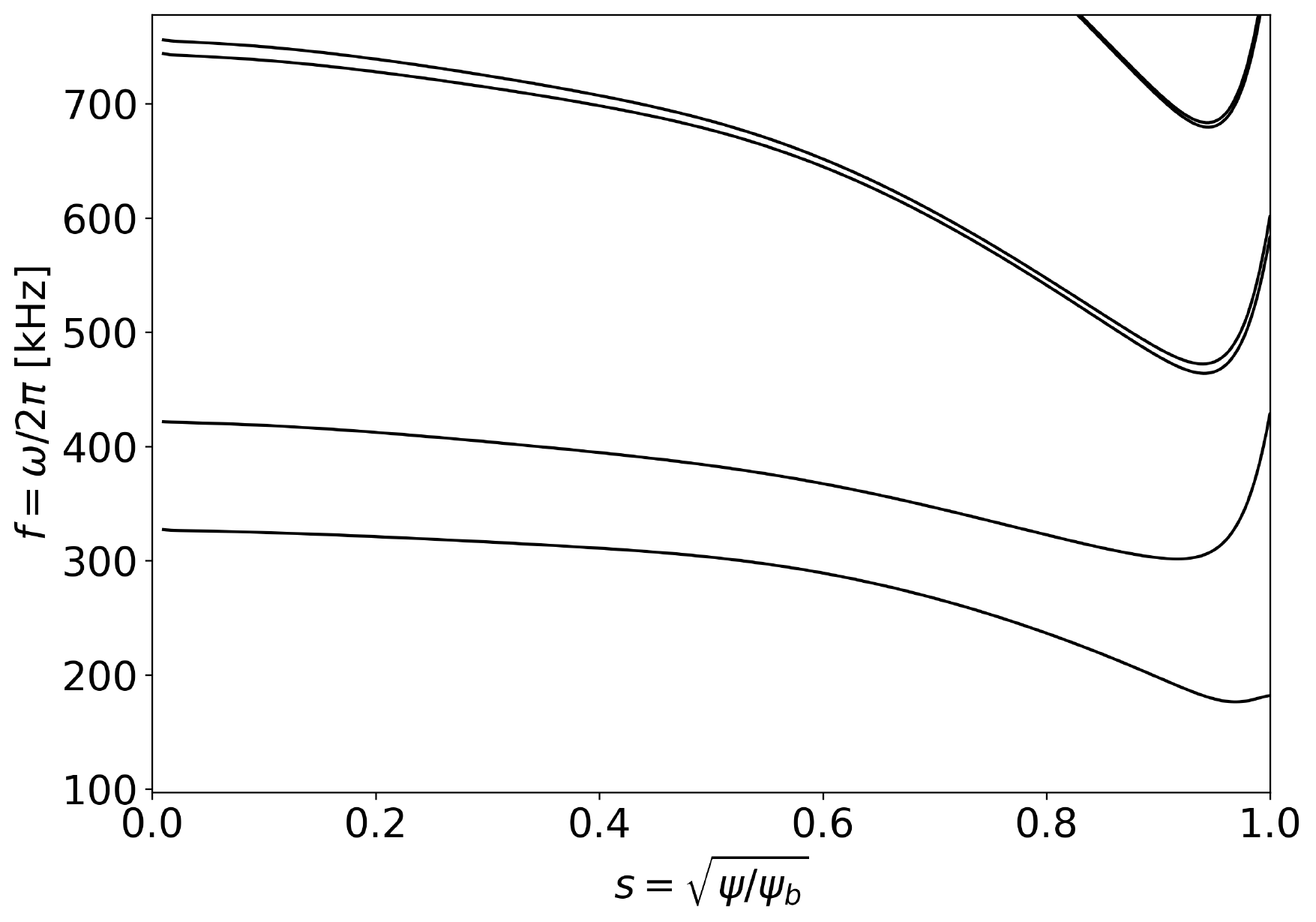}
	\caption{Axisymmetric shear-Alfvén continuum computed with \texttt{CSMISH} for pulse \#82221 at $t=8.07$ s, with the returned normalised frequency multiplied by $v_A(0)/2\pi R_0\approx390$ kHz. The minima of the first and second branches are reached at $s=0.97$ and $s=0.92$, respectively.}
	\label{fig:SAC_82221_n0_4807}
\end{figure}

The magnetic equilibrium was reconstructed by the equilibrium code \texttt{EFIT} \cite{EFIT} with pressure constraints and refined by the Grad-Shafranov solver \texttt{HELENA} \cite{Huysmans1991}. The plasma mass density was defined from measurements of the electron density via High Resolution Thomson Scattering (HRTS), assuming the condition of quasi-neutrality ($n_e = n_i$). To model the electron density profile, $n_e(s)$, in terms of the radial coordinate $s = \sqrt{\Psi/\Psi_b}$ ($ \approx r/a$), a function defined by the product of a third-order polynomial and a hyperbolic tangent was fitted to the HRTS data, as shown in figure \ref{fig:ne_q_t4807}. The $n=0$ shear-Alfvén continuum computed by \texttt{CSMISH} for this equilibrium is plotted in figure \ref{fig:SAC_82221_n0_4807}. At the chosen time instant, the plasma is in an evolving Ohmic-heated L-mode; hence, the pedestal of the density profile is not as steep as in later phases. This results in a smoother SAC, with wider angled branches at the minima.

\begin{figure}[t]
	\centering
	\includegraphics[width=\linewidth]{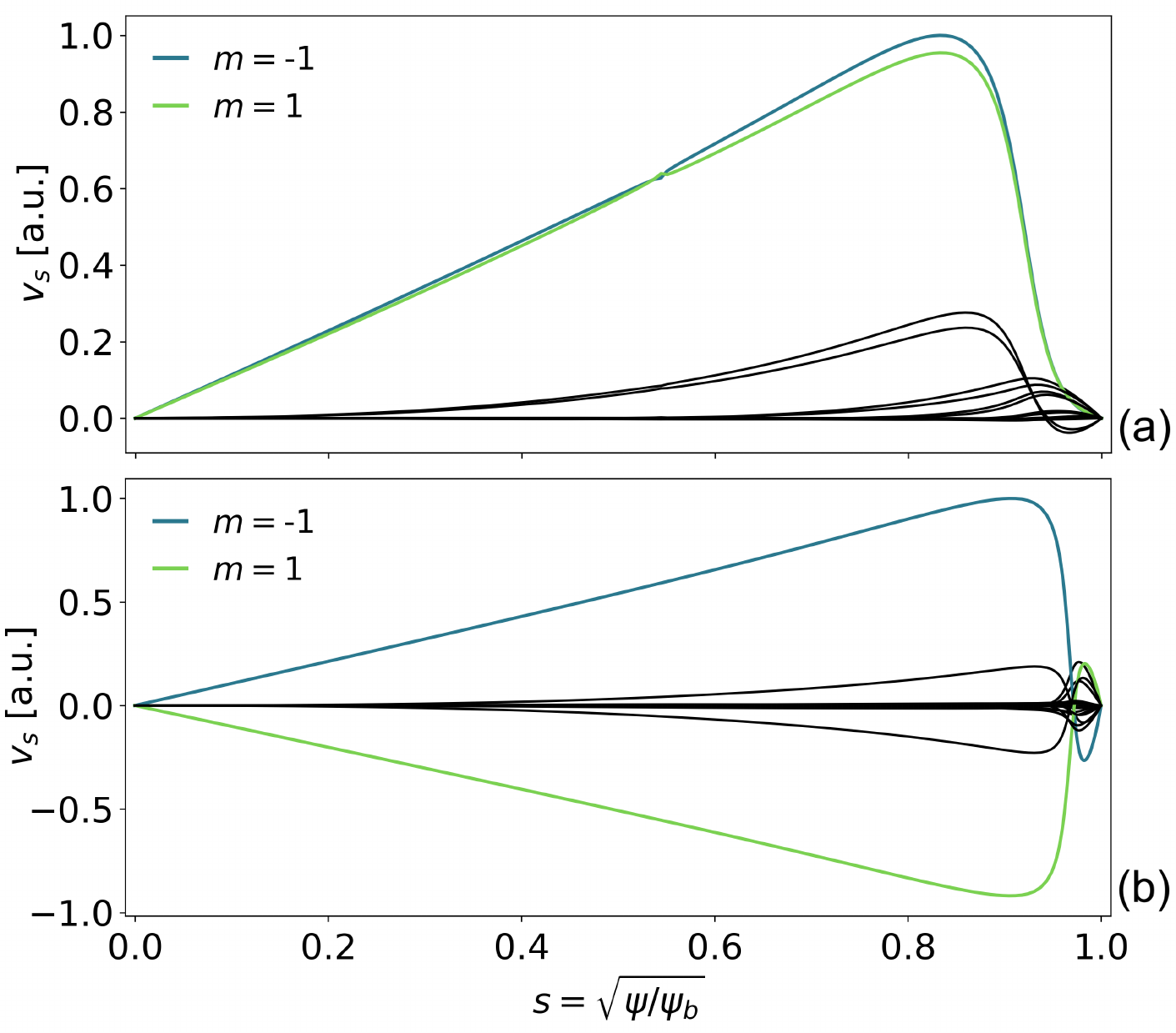}
	\caption{Radial component of the perturbed velocity, $v_s$, of the GAEs found by \texttt{MISHKA} under the (a) second and (b) first minimum of the SAC for pulse \#82221 at $t=8.07$ s.}
	\label{fig:radial_structure_modes}
\end{figure}

Using the ideal MHD code \texttt{MISHKA} \cite{Mikhailovskii1997}, we searched for discrete AEs lying below the first and second $n=0$ SAC minima, reached at $f_\text{SAC-1}=176$ kHz and $f_\text{SAC-2}=302$ kHz, respectively. A single GAE was found under each of them, with eigenfrequencies $f_\text{GAE-1} = 175$ kHz and  $f_\text{GAE-2} = 298$ kHz, i.e. very close to the values of the SAC minima. This small difference was consistently obtained for multiple study cases. The radial structures of the first and second GAEs are dominated by the $m=\pm1$ poloidal harmonics, as depicted in figure \ref{fig:radial_structure_modes}. The GAE-1 has odd parity (i.e., antibalooning mode structure, with symmetric Fourier harmonics having opposite sign), while the GAE-2 has even parity (or a balooning structure). The largest gradient of the radial component of the perturbed velocity is reached precisely around the radial location of the corresponding minimum. Although the GAE-2 intersects the continuum around $s \approx 0.54$, the singularity in the mode structure is small such that, if the driving factor is sufficiently strong and the continuum damping rate low enough, as proven in earlier publications \cite{Villard1997}, the mode can still be driven unstable. Both GAEs were also found using ideal MHD code \texttt{CASTOR} \cite{Kerner1998} in the compressible regime, confirming their weak coupling with the acoustic continuum and their incompressible shear-Alfvén nature.

For this particular case study, the experimental frequency of the HFO-2 matches that of the numerical GAE-2, whereas the frequency of the HFO-1 has a discrepancy of $\sim 20\%$ relative to that of the GAE-1 (and respective SAC minimum). This may eventually come from uncertainties in the estimation of the denormalisation factor $v_A(0)/2\pi R_0$ (e.g. $n_e(s)$ is not measured experimentally for $s < 0.2$, hence $n_e(s=0)$ was obtained by extrapolation) combined with an unreliable equilibrium reconstruction near the edge, which determines the SAC shape and gap size. In addition, it is conceivable that the lower branch is more sensitive to variations in equilibrium parameters than the upper one. If so, a slight inaccuracy in a certain parameter could alter the lower branch without significantly affecting the upper one. This would justify the agreement between $f_\text{SAC-2}$/$f_\text{GAE-2}$ and $f_\text{HFO-2}$, along with the discrepancy between $f_\text{SAC-1}$/$f_\text{GAE-1}$ and $f_\text{HFO-1}$. Nonetheless, this deviation does not change the main result of the analysis presented here: the numerical GAEs frequencies can be estimated with high accuracy (relative error $\lesssim 2 \%$) from the SAC minima. This finding will serve as the foundation for the method employed in the next section, where we extend the comparison between theoretical and experimental results to a statistically meaningful sample.

\subsection{Extended analysis: a predictive model for HFOs frequencies}\label{sec:Extended_analysis}

Motivated by the fact that GAEs tend to lie just below the continuum minima, the approach adopted here to show that the frequencies of HFOs can be reliably predicted by a simple ideal MHD model was to plot the experimental frequencies of the $n=0$ HFOs, measured in the spectrogram of $\dot{B}_\theta$, against the numerical frequencies of the corresponding $n=0$ SAC minima, computed by \texttt{CSMISH} from a suitably reconstructed equilibrium, for a comprehensive set of pulses. Overall, a total of 14 pulses and 40 time instants were used in this analysis, covering a wide set of plasma compositions (H$^1$, D, D-He$^3$, D-T) and parameters ($I_p= 1.2-3.0$ MA, $B=1.8-3.65$~T, $n_e(0)=2.4-6.8$ m$^{-3}$) that display up to three HFO bands in the spectrogram of the magnetic signal. Given the considerable number of time instants under analysis, it must be stressed that the method employed saves a significant amount of computational effort, as solving simultaneously for the frequency and radial structure of the AE is far more expensive than evaluating SAC frequencies within a small radial range near the pedestal. Moreover, estimating GAEs' frequencies from the SAC minima is a more robust and practical approach than the actual eigenmode calculation.

It is worth mentioning that the equilibrium reconstructed using \texttt{EFIT} with external magnetic constraints alone, pressure constraints (\texttt{EFTP}), and MSE constraints (\texttt{EFTM}) can vary significantly, leading to discrepancies between both the shape and value (around $\sim 15$ kHz) of the continuum minima computed for each equilibrium. Although \texttt{EFTM} yields a more reliable $q$ profile, which is determinant for the accuracy of the SAC, MSE data is only available for a few pulses of interest (here, only \#86775 and \#86811). Thus, most of our analysis was carried out using pressure-constrained equilibrium reconstruction.

\begin{figure}[t!]
	\centering
	\includegraphics[width=\linewidth]{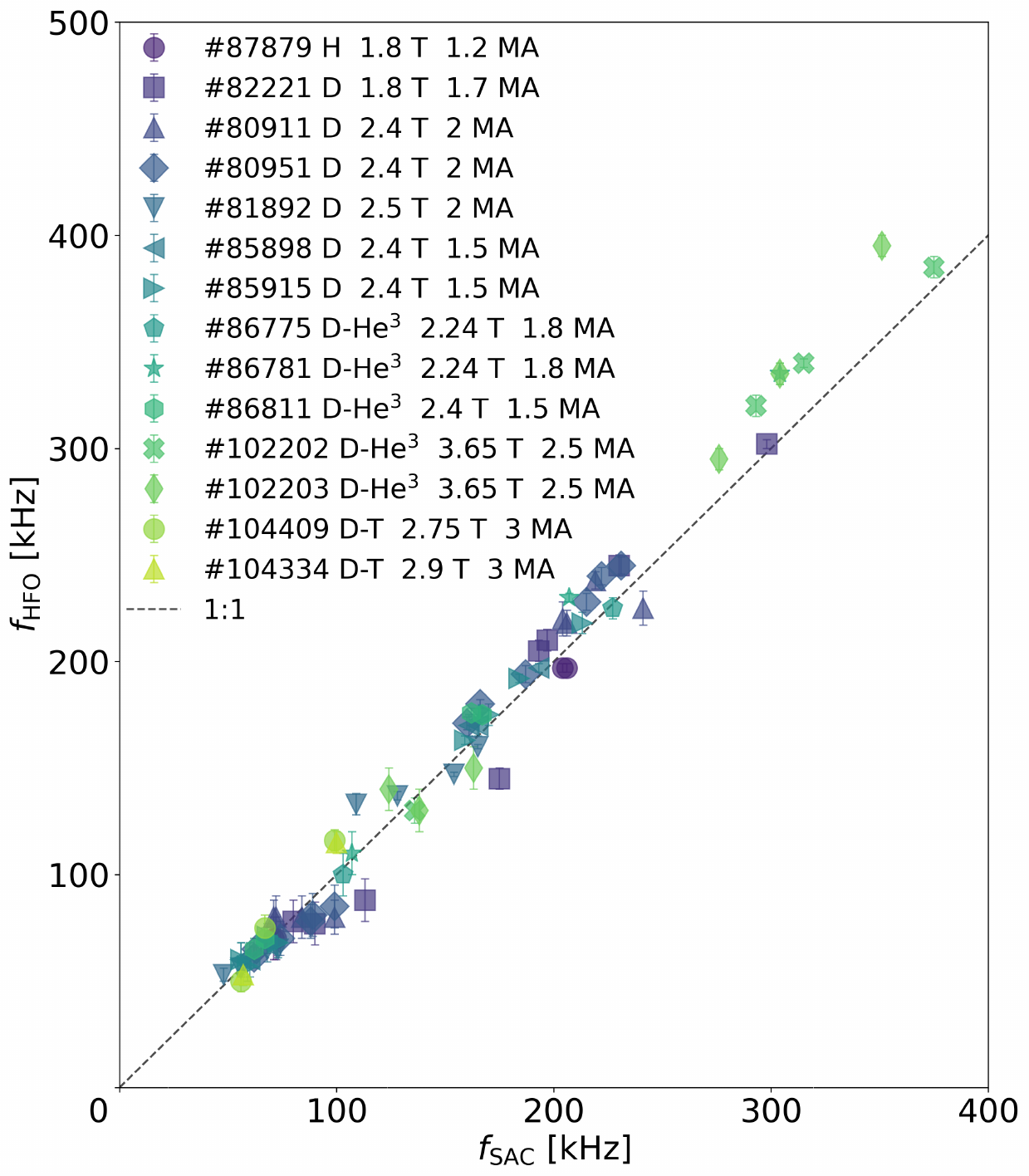}
	\caption{Plot of the measured $n=0$ HFOs frequencies, $f_\text{HFO}$, versus the frequency of the corresponding $n=0$ shear-Alfvén continuum minima computed by \texttt{CSMISH}, $f_\texttt{SAC}$.}
	\label{fig:CSMISH_vs_exp_frequencies}
\end{figure}

The results obtained are plotted in figure \ref{fig:CSMISH_vs_exp_frequencies} and reveal a remarkable agreement between experimental observations and numerical results, thereby corroborating the hypothesis that HFOs are GAEs. As expected, the agreement improves when \texttt{EFTM} is used (relative error $<5\%$, for both modes). This highlights the potential of using the measured frequencies of $n=0$ HFOs as MHD markers, placing additional constraints at the plasma pedestal for reliable equilibrium reconstruction.

This conclusion is in line with the frequency modulation of HFOs, which is mainly dictated by the variation of $1/\sqrt{\rho}$ at the pedestal, as depicted in figure \ref{fig:discharge_80951_plots_inv_sqrt_LID4_mult_factor}(a). Naturally, it also depends on $q$ and $\kappa_s$, but, as can be seen in figure \ref{fig:discharge_80951_plots_inv_sqrt_LID4_mult_factor}(b), the variation of the edge plasma elongation and $q$ over time is low for $t>12$ s.

\subsection{Non-ideal MHD contributions and mode calculation}\label{sec:Resistivity}

\begin{figure}[t!]
	\centering
	\includegraphics[width=\linewidth]{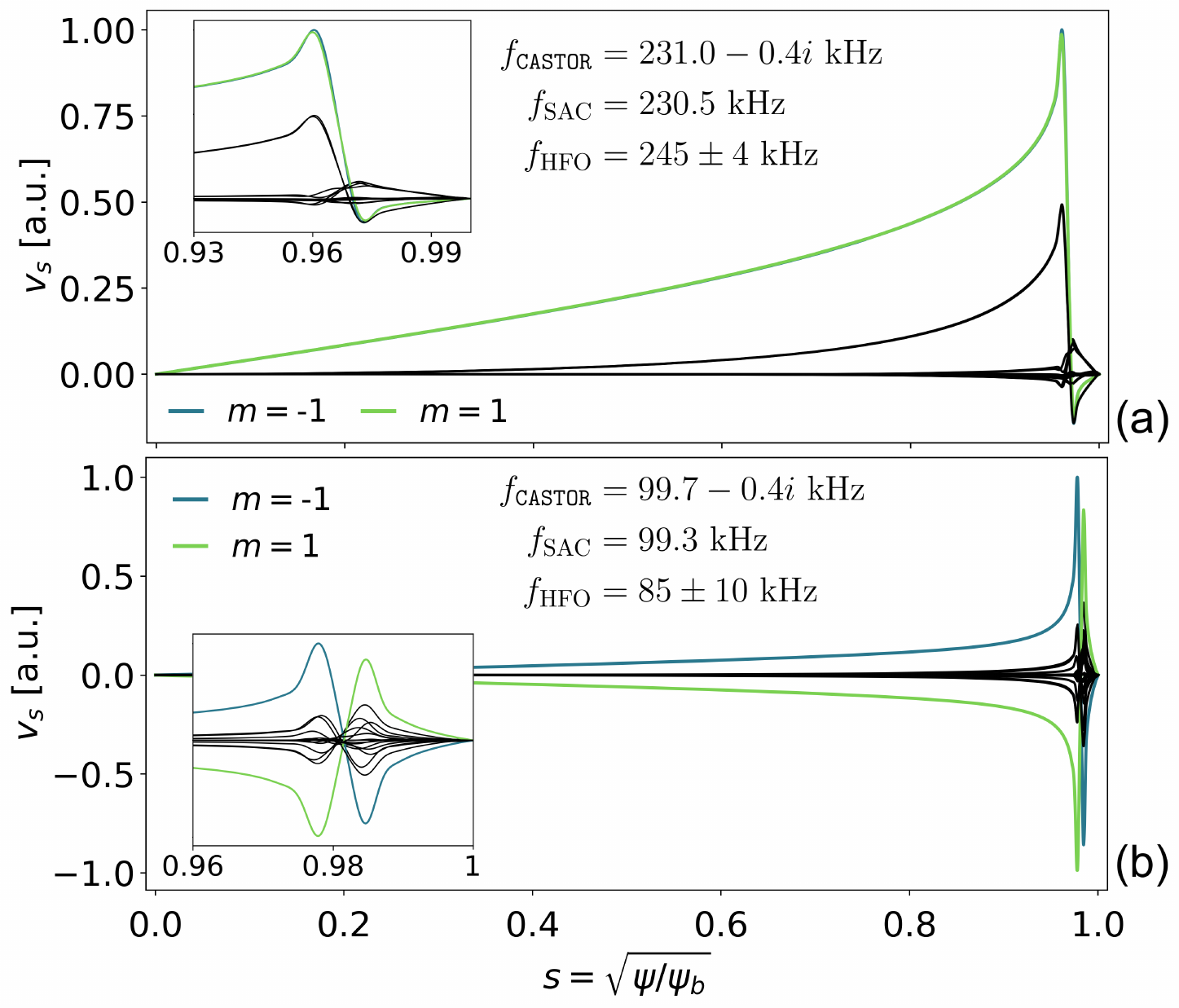}
	\caption{Radial component of the perturbed velocity, $v_s$, of the eigenfunctions found by \texttt{CASTOR} under incompressibility and finite resistivity near the (a) second and (b) first minimum of the ideal MHD SAC for pulse \#80951 at $t=11.77$ s.}
	\label{fig:radial_structure_modes_res}
\end{figure}

Even though the agreement conveyed by figure \ref{fig:CSMISH_vs_exp_frequencies} shows unequivocally that ideal MHD can accurately predict the frequencies of HFOs, non-ideal contributions may be relevant to the numerical calculation of the modes. Here we do not intend to explore this question in depth, but merely provide a brief insight into the reasons behind this statement, as well as advance possible interpretations.

While for a set of cases \texttt{MISHKA} computed an AE with frequency close to the second minimum of the continuum, it generally failed to converge to a mode below the first minimum. Two factors may be contributing to this behaviour. The first is the eventual low accuracy of the equilibrium reconstruction at the pedestal and close to the plasma boundary, as the existence of GAEs was shown to depend primarily on the $q$ and $B_\phi$ profiles \cite{Villard1997}. Near the edge, the reconstructed $q$ profile, along with other plasma parameters, is subjected to uncertainties (as the constraints used are conditioned by the spatial resolution of the diagnostics and the data measured at the edge are subjected to large uncertainties), which may then reflect on the successful or unsuccessful numerical calculation of the modes for the equilibrium in question.

\begin{figure}[t!]
	\centering
	\includegraphics[width=\linewidth]{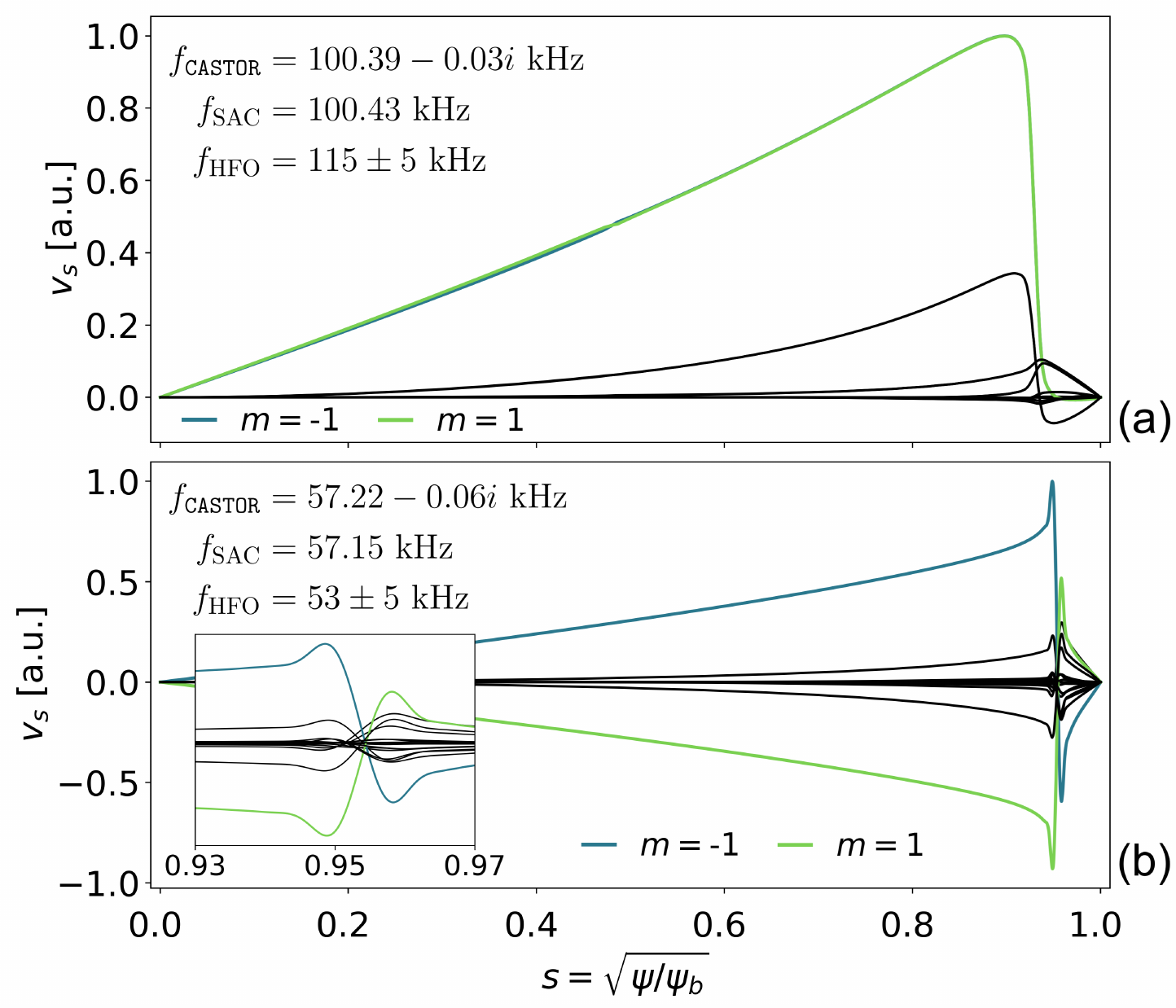}
	\caption{Radial component of the perturbed velocity, $v_s$, of the eigenfunctions found by \texttt{CASTOR} under incompressibility and finite resistivity near the (a) second and (b) first minimum of the ideal MHD SAC for pulse \#104334 at $t=5.62$ s.}
	\label{fig:radial_structure_modes_res_104334}
\end{figure}

The second factor concerns the limitations of ideal MHD theory. Numerically, we found that as the pedestal grows steeper and the SAC sharpens around the minima, the frequency of the computed GAE shifts upwards, eventually stepping into the continuum, where \texttt{MISHKA} converges to singular solutions. In these conditions, a more accurate description of these AEs would require keeping non-ideal terms in the set of MHD equations to lift the singularity of the eigenfunctions. Since the plasma temperature near the edge is low  ($< 1$ keV), it is not unreasonable to assume that the resistivity may have a non-negligible contribution. To check this hypothesis, we chose five time instants for which the ideal-MHD code \texttt{MISHKA} did not converge properly under the first minimum (\#80951, $t=11.77$, $24.52$, $25.82$ s; \#86781, $t=14.50$ s; \#104334, $t=5.62$ s) and employed the resistive MHD code \texttt{CASTOR} instead, keeping incompressibility and adding a finite resistivity $\eta$ corresponding to an inverse magnetic Reynolds number $\eta/(\mu_0 v_A(0) R_0) \sim 10^{-9}$. Indeed, \texttt{CASTOR} converged to continuous and non-singular global eigenfunctions with complex eigenfrequencies, whose real part closely follows the SAC from ideal MHD theory (already within it, as expected), and a non-negligible imaginary part ($\mathfrak{Im}(\omega)\sim 1-0.1$ kHz). By comparing the structure of the GAEs obtained for different equilibria, we observed that, as the frequency of the resistive mode enters the SAC, the radial component of the perturbed velocity, $v_s$, narrows around the maximal amplitude, as illustrated in figures \ref{fig:radial_structure_modes_res} and \ref{fig:radial_structure_modes_res_104334}. Conversely, as the frequency approaches the minimum from above, $v_s$ widens and tends toward the mode structures of figure \ref{fig:radial_structure_modes}.

\begin{figure*}[t!]
	\centering
	\includegraphics[width=0.85\linewidth]{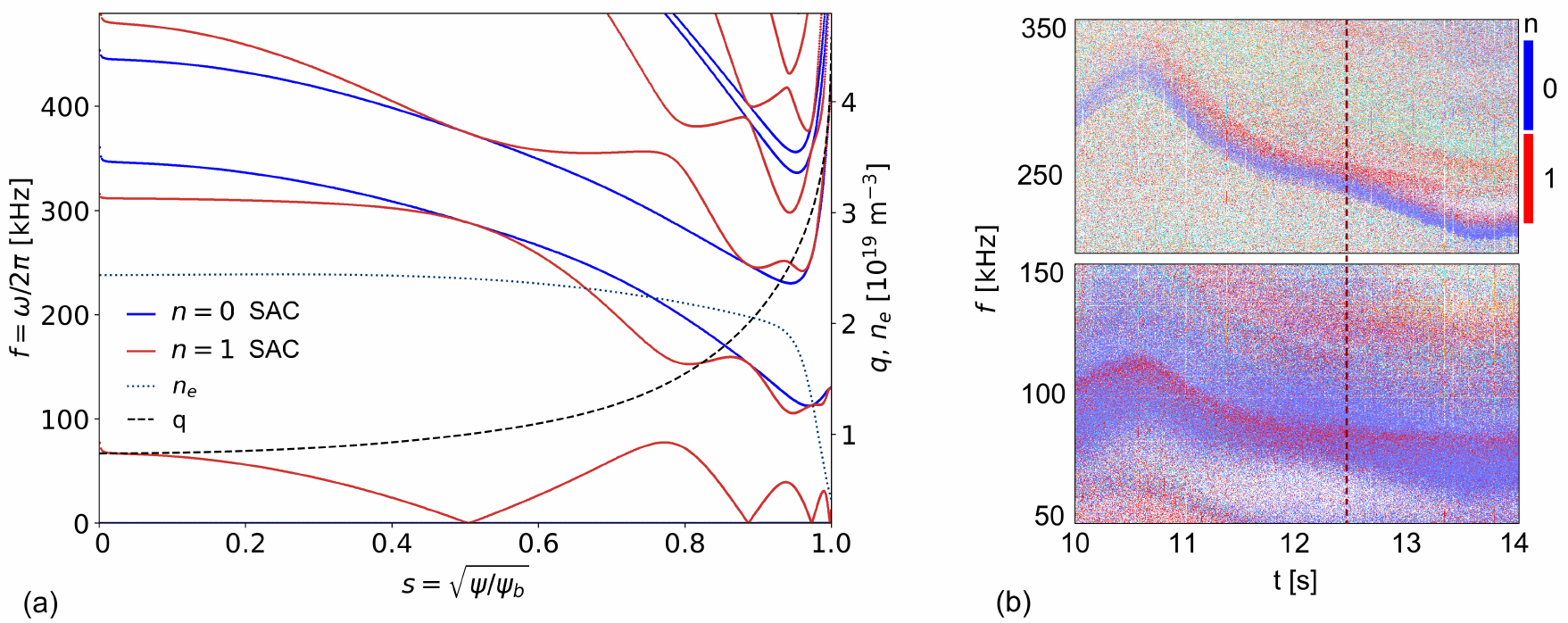}
	\caption{(a) Plot of the $n=0$ and $n=1$ shear-Alfvén continuum computed by \texttt{CSMISH}, with the returned normalised frequency multiplied by $v_A(0)/2\pi R_0 \approx 326$ kHz, along with $n_e$ and $q$ profiles computed for \#82221 at $t=12.47$ s; (b) spectrogram displaying the toroidal mode number for the time window $10-14$ s, where both $n=0$ (blue) and $n=1$ (red) HFOs bands are visible. The dotted line corresponds to the time instant for which the continua were computed.}
	\label{fig:SAC_t5247_n0_n1_full_bar_blue}
\end{figure*}

It is important to note that, for the five time instants analysed here, the difference found between the real part of the eigenvalues and the respective continuum minima is $\lesssim 0.5$ kHz (or relative error $\lesssim 0.6$ \%). This result, along with those of the previous subsection, clearly shows that the frequencies of HFOs are closely reproduced by the ones of the SAC minima. Hence, while non-ideal effects may affect the AEs' radial structure, they do not significantly alter the frequency, and, for practical purposes, the latter can be accurately estimated using ideal MHD theory.

\section{HFOs with finite toroidal mode number}\label{sec:HFOs_of_finite_n}

As already mentioned in section \ref{sec:Experimental_Description_HFOs}, non-axisymmetric high-frequency oscillations modulated by their axisymmetric counterparts are observed in the spectrograms of figures \ref{fig:shot_80951} and \ref{fig:discharge_82221_plots_spec}.
We can typically distinguish two $n=1$ HFOs bands, the first appearing below or superposed to the $n=0$ HFO-1 and the second above the $n=0$ HFO-2.
We advance the hypothesis that these $n=1$ bands are $n=1$ AEs bound to the pedestal region where the SAC reaches a minimum.
Following the same logic as before, here we perform a simple comparison between the continuum minima and the frequencies of the bands to show that experimental observations are consistent with predictions of ideal MHD theory.

Although the profiles of the axisymmetric and non-axisymmetric continua are fundamentally different, it is generally observed that the $n=0$ continuum shapes the $n \neq 0$ continua, forming its envelope. Concretely, the coupled branches of the $n \neq 0$ SAC follow the ones of the axisymmetric SAC such that, if each branch of the latter reaches an absolute minimum at the edge, so will happen to the ones of the former. The frequencies of $n=0$ and $n\neq0$ SAC absolute minima at the edge are, hence, not expected to deviate significantly.

Figure \ref{fig:SAC_t5247_n0_n1_full_bar_blue} displays the $n=0$ and $n=1$ SAC for pulse \#82221 at $t=12.47$~s, where both lower and upper $n=1$ bands are visible in the toroidal mode number spectrogram.
The absolute minima reached by the two lowest branches of the $n=1$ SAC differ from the frequencies of the corresponding $n=0$ minima by $f_{ \text{SAC-1}}^{n=1}-f_{\text{SAC-1}}^{n=0} = -8$ kHz and $f_{\text{SAC-2}}^{n=1}-f_{\text{SAC-2}}^{n=0} = 12$ kHz, each being attained at similar radial positions. This numerical result is consistent with the frequency difference between the $n=0$ and $n=1$ bands measured experimentally, $f_{ \text{HFO-1}}^{n=1}-f_{\text{HFO-1}}^{n=0} \approx 0 \pm 10$ kHz and $f_{\text{HFO-2}}^{n=1}-f_{\text{HFO-2}}^{n=0} \approx 10 \pm5$ kHz, therefore supporting our hypothesis.
It should be noted that, for non-axisymmetric modes, the measured frequency is Doppler-shifted by the plasma rotation, which is generally inferred from charge exchange spectroscopy data. Yet, since the latter was not available for this pulse, we could not perform the required correction to get the mode frequency in the plasma reference frame. Nevertheless, in the absence of NBI, it is reasonable to assume that plasma rotation will be low enough ($\lesssim 5$ kHz) to be covered by the experimental uncertainty of the measured HFO frequency. Furthermore, the modes are tied to the plasma edge, where the plasma rotation is expected to be small.

\section{Summary and discussion}\label{sec:conclusions}

We demonstrated that the frequencies of axisymmetric HFOs measured at JET in a variety of plasma scenarios agree with the frequencies of the $n=0$ shear-Alfvén continuum minima computed numerically. This result shows that ideal MHD theory can accurately predict the frequencies of such modes and supports the hypothesis that $n=0$ HFOs are $n=0$ GAEs arising below the SAC minima produced by sharp gradients of the safety factor and plasma density near the edge. Although the main motivation for this work was provided by HFOs observations during the L-H transition, we presented experimental evidence showing that such activity is a more general phenomenon, encompassing different plasma compositions and heating schemes.

Despite the eventual role of non-ideal MHD corrections (e.g., plasma resistivity) required for the numerical evaluation of the GAEs radial structure, ideal MHD theory remains a fairly convenient approach: the minima of the $n=0$ shear-Alfvén continuum can be used as a proxy to predict the frequencies of HFOs, or, reversely, the measured frequencies can be used to predict the values and positions of the SAC minima. Indeed, in a pragmatic perspective, the nature of these axisymmetric HFOs, along with the fact that they are measured over long time windows and across a variety of plasma scenarios, makes them suitable MHD markers to improve the reliability of equilibrium reconstruction near the edge. This can be particularly useful when no internal constraints are available. The analytic expression derived here for the two lowest branches of the $n=0$ SAC constitutes a first step in this direction, by establishing how HFOs frequencies depend on equilibrium parameters such as the plasma density, safety factor and elongation.

This work did not address the factor responsible for driving these Alfvén modes unstable. Yet, we have shown that $n=0$ GAEs appear in various plasma scenarios, in particular, during Ohmic heated plasma phases, implying that they are not driven unstable by energetic particles via the conventional mechanisms. The same applies to the analogous $n \neq 0$ AEs, which are observed under no external heating as well. Previous efforts \cite{McClements2002, Kryzhanovskyy2024,Marchenko2020, Maraschek1997} demonstrated that high-frequency Alfvén Eigenmodes can be excited in the absence of energetic particles. Possible excitation mechanisms for such AEs have also been proposed based on correlated low-frequency MHD activity \cite{McClements2002} and on the interaction between edge turbulence and Alfvén waves \cite{Marchenko2020, Maraschek1997}. Although the models provide a plausible explanation, the mechanism behind Alfvén modes destabilisation in the absence of energetic particles remains an open question. 

The ease with which these modes are excited and persist in a broad range of scenarios suggests they are likely to arise in future reactors, such as ITER. Due to their low amplitude, HFOs are not expected to affect the machine's operation. Nevertheless, they can serve the practical purpose mentioned above and constrain the equilibrium reconstruction at the pedestal region, which is home to a wide variety of phenomena, thereby improving the accuracy of modelling conditions.

\begin{acknowledgements}

This work has been carried out within the framework of the EUROfusion Consortium, funded by the European Union via the Euratom Research and Training Programme (Grant Agreement No. 101052200—EUROfusion). Views and opinions expressed are however those of the authors only and do not necessarily reflect those of the European Union or the European Commission. Neither the European Union nor the European Commission can be held responsible for them. IPFN activities were supported by FCT—Fundação para a Ciência e Tecnologia, I.P. by project reference \href{https://doi.org/10.54499/UID/50010/2025}{UID/50010/2025}. This research was supported in part by the Spanish grant PID2021-127727OB-I00, funded by MICIU/AEI/10.13039/501100011033 and by ERDF/EU.
\end{acknowledgements}

\bibliography{apssamp}

\end{document}